\documentclass[aps,prl,twocolumn,showpacs,superscriptaddress,floatfix]{revtex4}
\pdfoutput=1

\usepackage{graphicx}
\usepackage{dcolumn}
\usepackage{bm}
\usepackage{amsfonts}
\usepackage{amsmath}
\usepackage{color}

\begin{document}

\title{Velocity-modulation control of electron-wave propagation in graphene}

\author{Arnaud Raoux}
\thanks{Permanent address: Formation Interuniversitaire de Physique,
D\'epartement de Physique de l'\'Ecole Normale Sup\'erieure, 24 rue Lhomond, 75231 Paris Cedex 05, France}
\affiliation{NEST-CNR-INFM and Scuola Normale Superiore, I-56126 Pisa, Italy}
\author{Marco Polini}
\email{m.polini@sns.it}
\affiliation{NEST-CNR-INFM and Scuola Normale Superiore, I-56126 Pisa, Italy}
\author{Reza Asgari}
\affiliation{School of Physics, Institute for Research in Fundamental Sciences (IPM), Tehran 19395-5531, Iran}
\author{A.R. Hamilton}
\affiliation{School of Physics, University of New South Wales, Sydney NSW 2052, Australia}
\author{Rosario Fazio}
\affiliation{NEST-CNR-INFM and Scuola Normale Superiore, I-56126 Pisa, Italy}
\author{A.H. MacDonald}
\affiliation{Department of Physics, University of Texas at Austin, Austin, Texas 78712, USA}

\begin{abstract}
Wave propagation control by spatial modulation of velocity has a long
history in optics and acoustics.
We address velocity-modulation control of electron wave
propagation in graphene and other two-dimensional Dirac-electron systems,
pointing out a key distinction of the Dirac-wave case.  We also propose a strategy for
pattern transfer from a remote metallic layer that is based on many-body
velocity renormalization.
\end{abstract}

\pacs{73.23.Ad,71.10.-w,78.67.Pt}

\maketitle

\noindent{\it Introduction}---
Control of electromagnetic and mechanical wave propagation by spatial modulation of wave velocity~\cite{photoniccrystals,phononiccrystals}
has been studied for many decades in optics and acoustics, originally in relatively simple
multi-layer structures which have a wealth of practical applications and more recently in sophisticated two-
and three-dimensional photonic and phononic crystal structures which can have gaps between transmission bands.

Recent advances~\cite{reviews} in the isolation and control of single and few-layer graphene
electron systems motivate a close examination of {\it velocity-modulation}
control in this material. Graphene is an allotrope of carbon atoms tightly packed in a two-dimensional  (2D)
honeycomb lattice. At energies near the Fermi energy of a neutral system, electron waves in graphene are described by a
2D massless Dirac equation and, like electromagnetic and mechanical waves, travel with a velocity that
is independent of wavelength. Because of this property, the analogies of electron transport with both optics~\cite{opticsanalogy}
and acoustics are stronger than in the conventional non-relativistic electron-wave case~\cite{leroyreview}.
In this article we first consider the propagation of massless Dirac fermion (MDF) waves through a medium with a position-dependent velocity, highlighting a key distinction between a Dirac wave and electromagnetic or acoustic waves.
We then discuss a non-invasive strategy for achieving velocity modulation in graphene without any direct physical contact to the sample, by transferring a spatial pattern from remote metal layers {\it via} many-body velocity renormalization.

\begin{figure}
\centering
\includegraphics[width=1.00\linewidth]{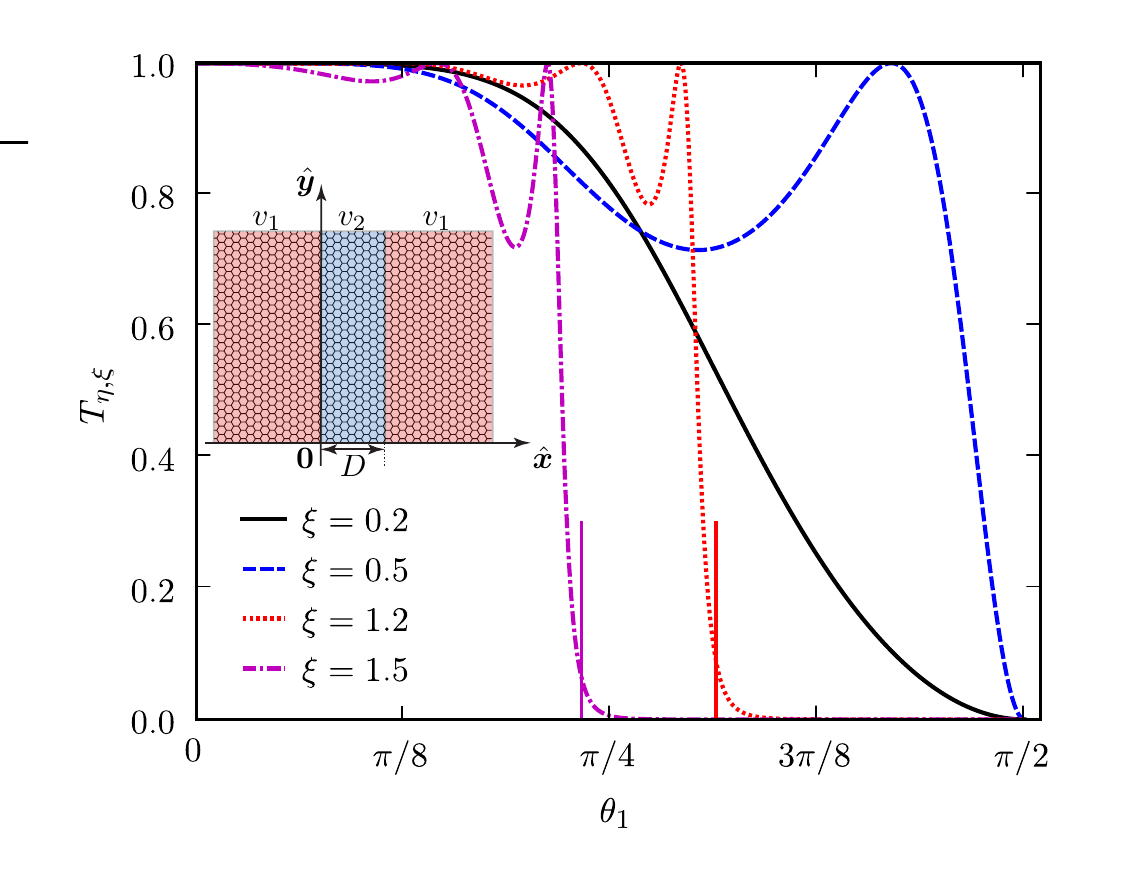}
\caption{(Color online) Inset: Cartoon of a velocity barrier. The  velocity of the massless carriers changes along the ${\hat {\bm x}}$ direction according to the simple functional form defined by Eq.~(\ref{eq:velocity_barrier}). Main panel: Transmission probability $T_{\eta,\xi}(\theta_1)$ as a function of angle of incidence $\theta_1$ for velocity ratio $\xi= v_2/v_1$ equal to $0.2$, $0.5$, $1.2$, and $1.5$.  The angle of incidence is defined so that $\theta_1=0$ corresponds to propagation perpendicular to the interface. These data have been obtained for $kD = \eta = 4\pi$ where $k$ is the incoming wavevector and $D$ is the barrier thickness. This value is typical~\cite{katsnelson} and corresponds for example to
$k = 2\pi/\lambda$ with $\lambda=50~{\rm nm}$ and $D = 100~{\rm nm}$. Note that for $\xi>1$,
the transmission probability drops rapidly toward zero when the critical incidence angles,
indicated by vertical bars, are exceeded. ($\theta_{\rm c} \approx 0.99~{\rm rad}$ for $\xi=1.2$ and $\approx 0.73~{\rm rad}$ for $\xi = 1.5$.)\label{fig:one}}
\end{figure}

\noindent{\it Scattering of MDFs against a velocity barrier}---
The influence of velocity variation on propagation is best illustrated by the
simplest example, transmission through a {\it velocity barrier}~\cite{concha_arXiv}, as illustrated
in the inset in Fig.~\ref{fig:one}. We first solve this scattering problem, highlighting a key distinction between a Dirac wave and
electromagnetic or acoustic waves. We consider MDFs in a medium
in which the Fermi velocity $v$ changes as a function of the 2D position ${\bm r}$: $v = v({\bm r})$. The massless Dirac equation in this case reads~\cite{peres_jpcm_2009}
\begin{equation}\label{eq:hermitean_Dirac}
{\hat {\cal H}} \Psi({\bm r})= -i \hbar \sqrt{v({\bm r})}\; {\bm \sigma} \cdot \nabla_{\bm r}\; [\sqrt{v({\bm r})}\Psi({\bm r})] = E~\Psi({\bm r})~,
\end{equation}
where $\Psi({\bm r}) =(\Psi_{\rm A}({\bm r}), \Psi_{\rm B}({\bm r}))^{\rm T}$ is a two-component spinor, $\Psi_{\rm A}({\bm r})$ and $\Psi_{\rm B}({\bm r})$ are
the honeycomb sublattice components
of the electron wave, and ${\bm \sigma} = (\sigma_x, \sigma_y)$ is a 2D Pauli matrix vector.  In using the Dirac equation
we are assuming that velocity variations are slow on a lattice constant scale.
In this limit spin and valley degrees of freedom play a passive role.
Note that the Hamiltonian in Eq.~(\ref{eq:hermitean_Dirac}) does not differ from its uniform system counterpart
${\hat {\cal H}}_{\rm MDF} = -i\hbar v {\bm \sigma}\cdot \nabla_{\bm r}$ merely by the replacement $v \to v({\bm r})$:
as pointed out by Peres in Ref.~\onlinecite{peres_jpcm_2009} this prescription would lead to a non-Hermitian operator.
It is nevertheless convenient to introduce the auxiliary spinor $\Phi({\bm r}) = \sqrt{v({\bm r})}\Psi({\bm r})$
which satisfies
\begin{equation}\label{eq:hermitean_Dirac_simpler}
 -i \hbar v({\bm r}) {\bm \sigma} \cdot \nabla_{\bm r}\Phi({\bm r}) = E~\Phi({\bm r})~.
\end{equation}
For the barrier problem illustrated in Fig.~\ref{fig:one} $v({\bm r}) = v(x)$ changes only along the ${\hat {\bm x}}$ direction and
momentum along the ${\hat {\bm y}}$ direction is a good quantum number.

We solve Eq.~(\ref{eq:hermitean_Dirac_simpler}) for a simple velocity barrier
\begin{equation}\label{eq:velocity_barrier}
v(x) =
\left\{
\begin{array}{ll}
v_1, & {\rm if}~x< 0\\
v_2, & {\rm if}~0< x< D\\
v_1, & {\rm if}~x>D
\end{array}
\right.~.
\end{equation}
This step-like model is justified when the Fermi wavelength $\lambda$ is much larger than the characteristic width
over which $v(x)$ changes, as discussed at length
for the potential-barrier case in Ref.~\onlinecite{katsnelson}.  For a given ${\hat {\bm y}}$-direction wavevector $k_y$ we are left with two coupled one-dimensional first-order differential equations:
\begin{equation}\label{eq:separate}
-i\hbar v(x)\left(\frac{d}{d x} \mp k_y\right) \phi_{{\rm A}({\rm B})}(x) = E~\phi_{{\rm B}({\rm A})}(x)~.
\end{equation}
where $ \Phi({\bm r}) = \phi(x) e^{i k_y y}$ and the spinor $\phi(x) = (\phi_{\rm A}(x), \phi_{\rm B}(x))^{\rm T}$.
The first order equations for the spinor components can [because $v(x)$ is piecewise constant] be
combined into a second order equation satisfied by both:
\begin{equation}\label{eq:single}
\left\{\frac{d^2}{d x^2}+\left[\frac{E}{\hbar v(x)}\right]^2 - k_y^2\right\}\phi_i(x) = 0~.
\end{equation}
The solutions of Eq.~(\ref{eq:single}) can be written as
\begin{equation}\label{eq:phiA}
\phi_{\rm A}(x)=\left\{
\begin{array}{ll}
(e^{i k_x x}+r e^{-i k_x x}), & {\rm if}~x<0 \\
(a e^{i q_x x} + b e^{-i q_x x}), & {\rm if}~0<x<D \\
t e^{i k_x x}, & {\rm if}~x>D \\
\end{array}
\right.~,
\end{equation}
where the constants $a$, $b$, $r$, and $t$ are to be determined,
\begin{equation}\label{eq:snell}
\left\{
\begin{array}{l}
{\displaystyle k_x = \sqrt{\left(\frac{E}{\hbar v_1}\right)^2- k_y^2} = k\cos(\theta_1)}\vspace{0.1 cm}\\
{\displaystyle q_x = \sqrt{\left(\frac{E}{\hbar v_2}\right)^2-k_y^2} = k\sqrt{\frac{1}{\xi^2}- \sin^2(\theta_1)} }
\end{array}
\right.~,
\end{equation}
$k = E/(\hbar v_1)$, $\xi = v_2/v_1$ is the velocity ratio, and $\theta_1$ is the angle of incidence, {\it i.e.} $\tan(\theta_1) = k_y / k_x$. The corresponding expression
for $\phi_{\rm B}(x)$ can be obtained from Eq.~(\ref{eq:phiA}) using Eq.~(\ref{eq:separate}).
In Eq.~(\ref{eq:snell}) we can identify $\theta_2 = \tan^{-1}(k_y / q_x)$ as the angle of refraction and
thereby obtain a quantum version of the famous {\it Snell-Descartes law} of geometrical optics:
\begin{equation}\label{eq:snell_descartes}
\frac{\sin(\theta_1)}{\sin(\theta_2)} = \frac{v_1}{v_2} = \frac{1}{\xi}~.
\end{equation}
For $\xi>1$ and $\theta_1 > \theta_{\rm c} = \sin^{-1}(1/\xi)$, the classical total internal reflection angle,
$q_x$ is imaginary and we expect negligible transmission through thick barriers.
In this case the classical correspondence fails and the refraction angle is not well defined.

Explicit evaluation of the four coefficients in Eq.~(\ref{eq:phiA})
requires matching conditions at the two interfaces which we obtain by the following
argument.  Dividing both sides of Eq.~(\ref{eq:hermitean_Dirac_simpler}) by $v(x)$ and
integrating across either interface implies
that the auxiliary spinor $\Phi$ is continuous, and therefore that the physical $\Psi$ satisfies the following matching conditions:
\begin{equation}\label{eq:interface-conditions}
\left\{
\begin{array}{l}
{\displaystyle \Psi(0^+,y) = \frac{1}{\sqrt{\xi}}~\Psi(0^-,y)} \vspace{0.2 cm}\\
{\displaystyle \Psi(D^+,y) = \sqrt{\xi}~\Psi(D^-,y)}
\end{array}
\right.~.
\end{equation}
These discontinuities in $\Psi$ guarantee that the divergence of the local current ${\bm J}({\bm r}) = v({\bm r}) \Psi^\dagger({\bm r}) {\bm \sigma} \Psi({\bm r})$
vanishes.  Using these matching conditions we are able to obtain an explicit expression for
the transmission~\cite{reflection} probability $T_{\eta, \xi}(\theta_1) =|t|^2$~:
\begin{equation}\label{eq:transmission}
T_{\eta, \xi}(\theta_1) = \frac{\cos^2{(\theta_1)}[1-\xi^2\sin^2{(\theta_1)}]}{C_{\eta,\xi}(\theta_1)}~,
\end{equation}
where $\eta = kD$ and $C_{\eta, \xi}(\theta_1) = \cos^2{(\theta_1)}[1-\xi^2\sin^2{(\theta_1)}]
+ (1-\xi)^2\sin^2{(q_x\eta/k)}\sin^2{(\theta_1)}$.

\begin{figure}
\centering
\includegraphics[width=1.00\linewidth]{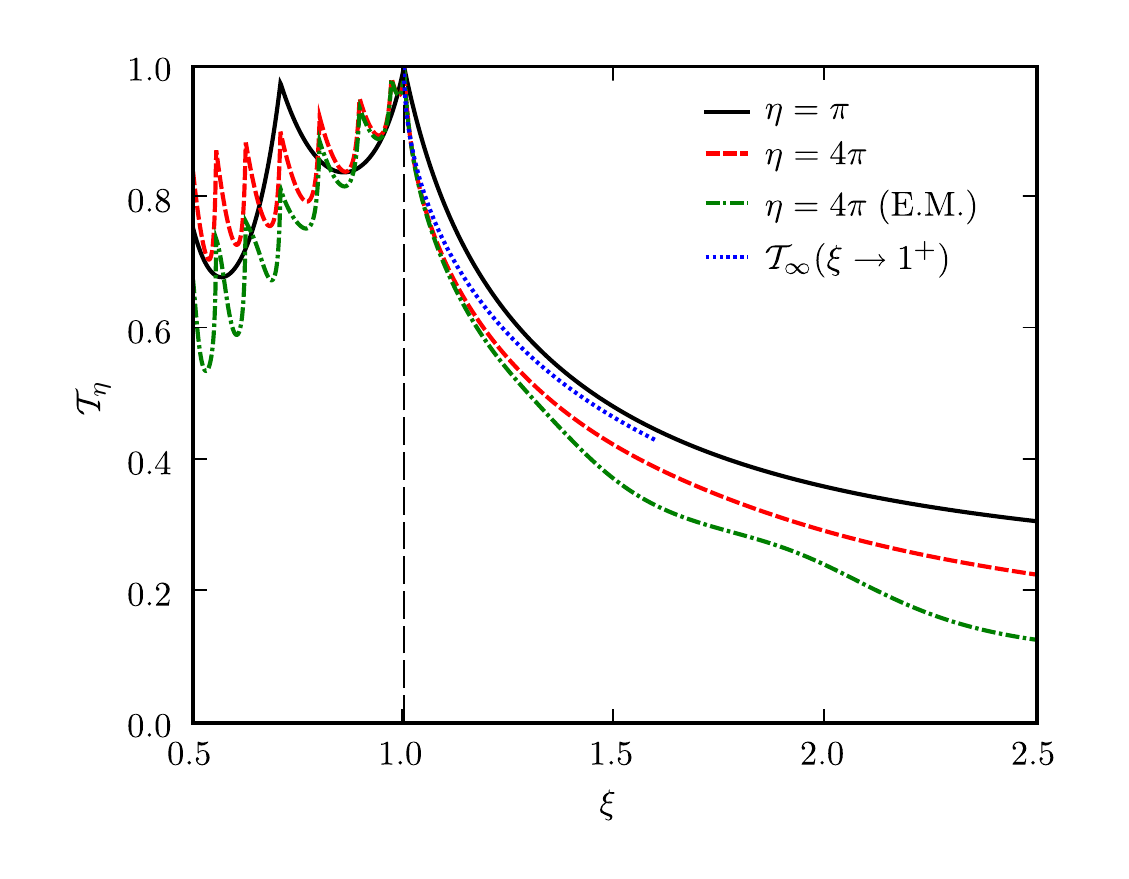}
\caption{(Color online) The integrated transmission ${\cal T}_\eta(\xi)$ as a function of the velocity ratio $\xi$ for $\eta=\pi$ [solid (black) line] and $\eta=4\pi$ [dashed (red) line]. The dotted (blue) line shows the classical limit ${\cal T}_{\infty}(\xi \to 1^+)$. The dash-dotted (green) line shows the integrated transmission for a linearly-polarized (along the ${\hat {\bm z}}$-direction) electromagnetic wave scattering against a non-magnetic barrier (in this case $\xi= \sqrt{\epsilon_1/\epsilon_2}$, where $\epsilon_i$ are the dielectric constants of the media composing the barrier, and $\eta = \omega D/c$). \label{fig:two}}
\end{figure}

In Fig.~\ref{fig:one} we plot the transmission probability $T_{\eta,\xi}$ as a function of $\theta_1$ for
$\eta = 4 \pi$ at several different values of the velocity ratio $\xi$.  Note that a velocity barrier is always perfectly transparent, $T_{\eta,\xi} \equiv 1$, for normal incidence ($\theta_1=0$) as in the standard Klein problem~\cite{katsnelson}.
This property of Dirac-wave propagation through a velocity barrier establishes a qualitative
difference between the present case and the familiar electromagnetic and acoustic cases,
and opens up new ground for the invention of spatial patterns with desirable
transmission properties.  It is also an important addition to the obvious
difference in velocity-wavelength relationship in distinguishing Dirac-wave
propagation from Schr\"{o}dinger-wave propagation.
In Fig.~\ref{fig:two} we plot the integrated transmission,
\begin{equation}
{\cal T}_\eta(\xi) = \frac{2}{\pi} \int_{0}^{\pi/2}d\theta_1~T_{\eta,\xi}(\theta_1)~,
\end{equation}
as a function of the velocity ratio $\xi$.
A sharp change in behavior is visible at $\xi = 1$
which foreshadows the total internal reflection properties of classical waves.
Indeed, in the $\eta \to \infty$ limit it is easy to prove that for $\xi = 1^{+}$ ${\cal T}_\eta(\xi) \to 2\theta_{\rm c}/\pi
\to 1 - 2\sqrt{2} \sqrt{\xi-1}/\pi$, which is a non-analytic function of $\xi$.
For this reason even a slight mismatch in velocities
can produce a large electron transport signal.

\noindent{\it Molding electron flow}--- 
Doped or gated graphene sheets are normal Fermi liquids~\cite{polini_arpes_2008,hwang_arpes_2008} but have a 
number of unusual quantitative features~\cite{polini_prl_2007} in their correlation physics which might provide an 
attractive route to velocity modulation, as we now explain. We first consider a grounded metal plane placed close to a graphene sheet.
The presence of the metal does not directly shift the chemical potential, like a biased gate would,
but because electron-electron interactions between MDFs are expected to be substantially screened and thus weaker,
has a rather large impact~\cite{alex} on the renormalized Fermi velocity of the quasiparticles close to the Fermi energy
that are important in transport.  We show below that quasiparticles under the screening plane move at
a speed $v^\star$ that is smaller than in an isolated graphene sheet~\cite{polini_ssc_2007,andrei_prl_2009}.
Using a single ground metal plane located close to a graphene sheet will thus lead to a very simple realization of the velocity barrier illustrated in the inset in Fig.~\ref{fig:one} with $\xi <1$. To realize a velocity barrier with $\xi>1$ one instead needs to use two metal gates located on top of the regions with $x<0$ and $x>D$. The area on top of the strip $0<x<D$ must instead be left empty.
In order for our abrupt-interface velocity-barrier calculation to be relevant,
the distance to the metal gate $d$ would have to be smaller than the Fermi wavelength $\lambda$, but
transmission properties will be similar even if this condition is not satisfied.
Any shape of velocity modulation can be achieved by transferring a suitable spatial pattern from the remote
lithographically-designed metal layer.

The effect of a metal gate on the quasiparticle velocity $v^\star$ can be estimated quantitatively by evaluating the
quasiparticle self-energy $\Sigma$ of an interacting MDF system near the quasiparticle pole.
We have generalized the GW theory~\cite{Giuliani_and_Vignale} calculations described in Ref.~\onlinecite{polini_ssc_2007}, replacing the bare Coulomb interaction with the corresponding expression appropriate
for a 2D electron system close to a perfect metallic screening plane: $V_d(q) = 2\pi e^2[1 - \exp(-2 q d)]/q$~\cite{electrostatics}.

\begin{figure}
\centering
\includegraphics[width=1.00\linewidth]{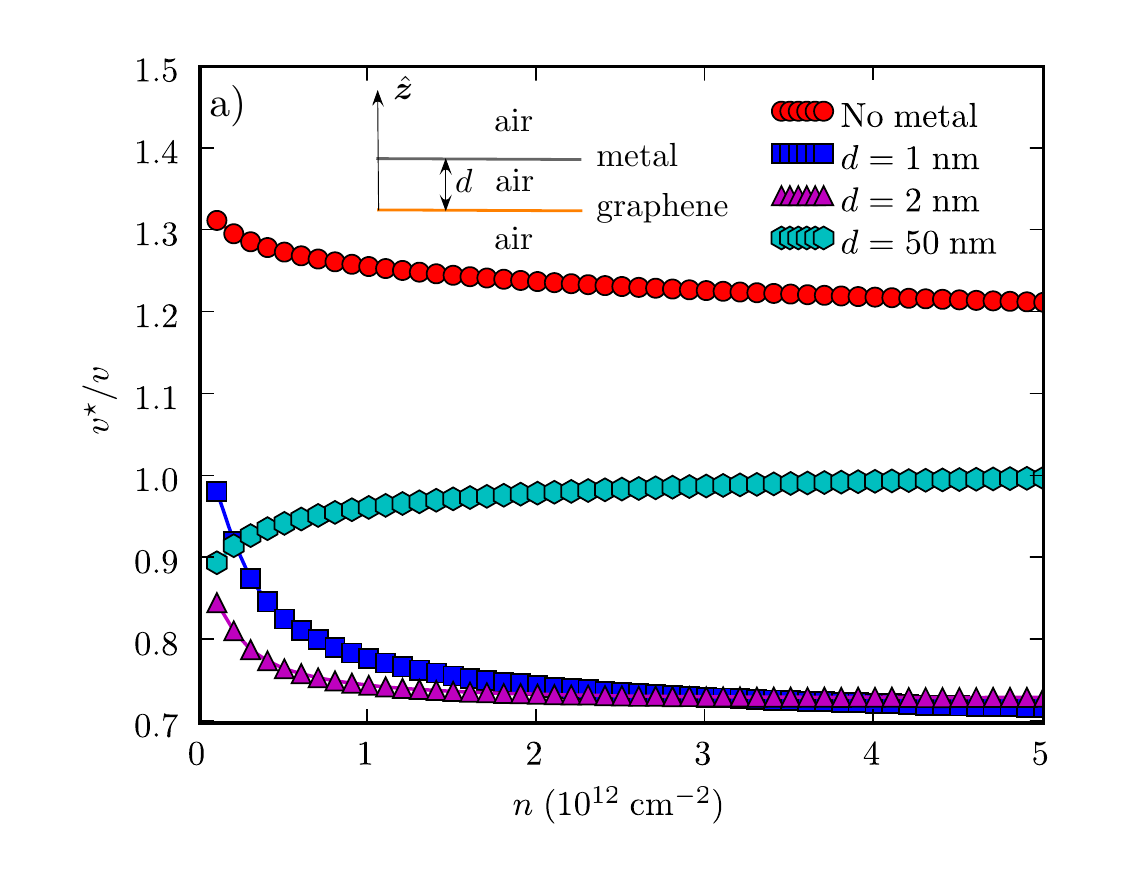}
\includegraphics[width=1.00\linewidth]{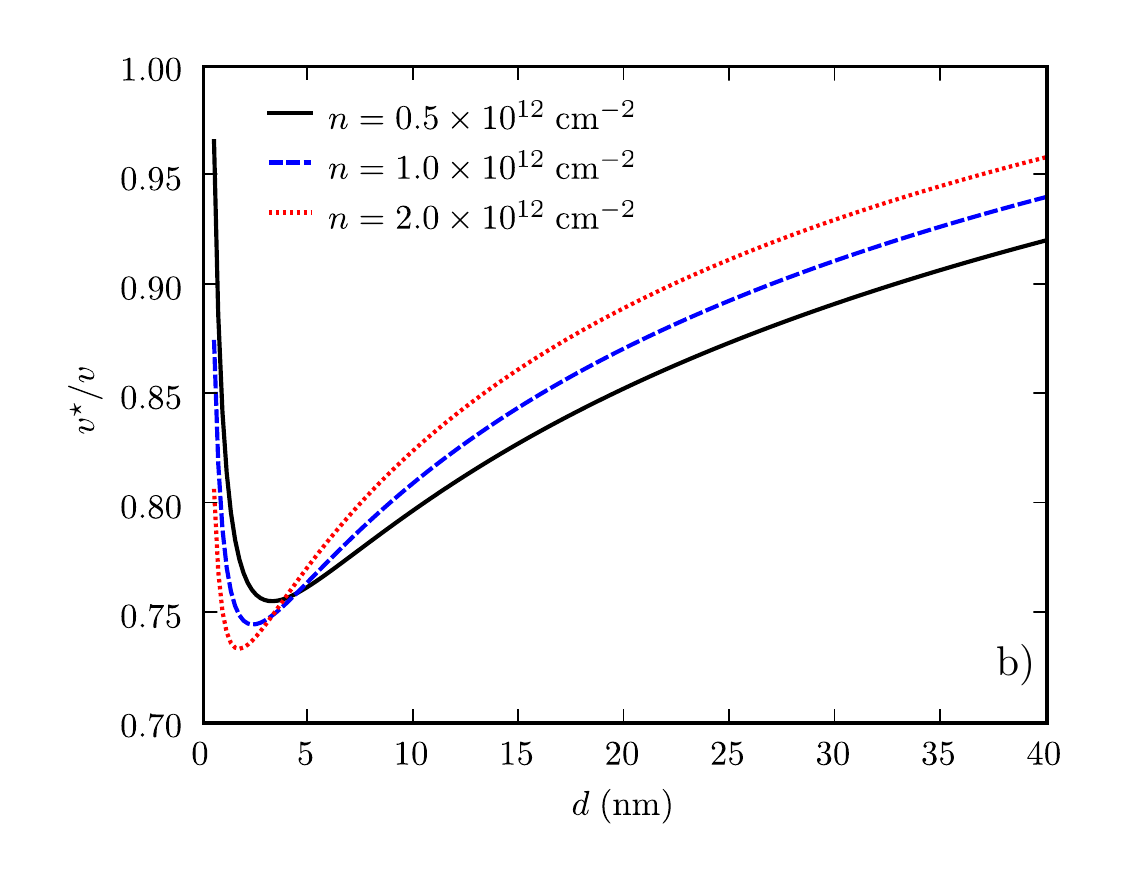}
\caption{(Color online) Panel a) Renormalized quasiparticle velocity $v^\star$ (in units of the bare velocity $v$) in graphene as a function of the electron density $n$ (in units of $10^{12}~{\rm cm}^{-2}$) in the presence of a ground perfect-metal plane located at a distance $d$ from the sheet (see inset). The (red) filled circles refer to an isolated sheet (in the absence of the perfect-metal plane). The other data refer to finite and increasingly larger (from top to bottom) values of $d$. Note how $v^\star/v$ is largely affected by the presence of the ground metal plane. Panel b) shows $v^\star/v$ as a function of $d$ (in ${\rm nm}$) for different values
of $n$: $v^\star/v$ is largely suppressed even when the gate is quite far away from the graphene sheet.\label{fig:three}}
\end{figure}

In Fig.~\ref{fig:three}a) we report numerical results for $v^\star/v$ as a function of electron density $n$
for several values of $d$, and in Fig.~\ref{fig:three}b) as a function of $d$ for fixed density.  We see that a substantial
velocity contrast can be induced by metallic gates that are tens of ${\rm nm}$'s from the graphene plane. The effect of the gate can extend much further if it is separated from the graphene by a dielectric with $\epsilon_{\rm r} \gg 1$.

In addition to the velocity modulation the remote metallic layer will produce a shift in chemical potential. The reason lies in the
fact that  the Fermi energy quasiparticles whose transmission properties we study here satisfy, strictly speaking, a {\it Dyson
equation}~\cite{polini_ssc_2007}, not a single-particle massless Dirac equation.  The chemical potential shift acts on the quasiparticles
exactly like an external potential acts on a free particle.  These shifts are small however, and can be compensated by biasing
the patterned metallic plane although it might  be difficult to completely separate velocity patterning from potential
patterning in real experiments.

\noindent{\it Discussion}--- Most of the considerations outlined in this paper apply equally well to any system in which matter
waves satisfy a massless Dirac equation, for example to the surface states of topological insulators~\cite{topological}.
Recently it has been proposed~\cite{park_nanolett_2009,gibertini_prb_2009,flindt_nanolett_2005} that MDFs can also
be realized in any standard 2D electron gas (2DEG), when appropriately nanopatterned.
Similar proposals to realize the Dirac spectrum have been discussed in the contexts of ultracold atoms in optical lattices~\cite{coldatoms} and photonic crystals~\cite{photonic}. Velocity modulation can be realized in these systems as well: a miniband structure is imprinted on a 2DEG subjected to a long-wavelength periodic external potential ({\it i.e.} a lateral superlattice) with hexagonal symmetry. If suitable conditions are satisfied~\cite{gibertini_prb_2009}, isolated Dirac points described by simple MDF Hamiltonians can appear in this miniband structure. As shown in Ref.~\onlinecite{gibertini_prb_2009} the Fermi velocity in these systems is quite sensitive to the strength of the periodic potential.
By patterning the surface of a 2DEG in such a way to create three regions along a given direction in which the strength of the external periodic potential changes one can achieve a velocity barrier similar to the one sketched in the inset in Fig.~\ref{fig:one}.

In summary, we have calculated the transmission probability
of massless Dirac fermions through a model ``velocity barrier" and showed how electrons flowing through it
obey the Snell-Descartes law of optics.  We have also discussed
a practical strategy for achieving substantial velocity modulation without damaging the graphene by exploiting
the influence of a remote metallic layer on many-body renormalization of the quasiparticle velocity.

\noindent{\it Acknowledgements}--- A.R. and M.P. acknowledge useful conversations with Diego Rainis. 
M.P. and A.R.H. acknowledge support from the Gordon Godfrey bequest and the ARC APF scheme respectively. 
A.H.M. acknowledges support from SWAN and the NSF-NRI program.


\begin{thebibliography}{77}
\bibitem{reflection}
	The reflection coefficient $r$ in Eq.~(\ref{eq:phiA}) is given by
	$
	r_{\eta,\xi}(\theta_1)= (\xi-1) \exp{(i\theta_1)}\sin(q_x \eta/k) \sin{(\theta_1)}/D_{\eta,\xi}(\theta_1)
	$,
	with $D_{\eta,\xi}(\theta_1) = (\xi q_x/k)\cos(q_x \eta /k)\cos{(\theta_1)}
	- i \sin(q_x \eta/k)[1-\xi\sin^2{(\theta_1)}]$.
\end{thebibliography}
\end{document}